\documentclass[pra,showpacs]{revtex4}

\usepackage{epsfig}

\usepackage{amsfonts}

\begin{document}

\title{Entanglement in thermal equilibrium states}
\author{O. Osenda}
\email{osenda@famaf.unc.edu.ar}
\author{G.A. Raggio}
\email{raggio@famaf.unc.edu.ar}
\affiliation{FaMAF-UNC, C\'ordoba, Argentina}
\begin{abstract}
 We revisit the problem of entanglement of thermal equilibrium states of composite systems. We introduce  characteristic, viz. critical, temperatures \textendash and bounds for them marking transitions from entanglement to separability. We present examples for the various possible thermal entanglement scenarios in bipartite qubit/qubit and qubit/qutrit systems.
\end{abstract}

\maketitle
\section{Introduction and some general results}

Consider a finite composite quantum system described by the complex  Hilbert space ${\cal H}={\cal H}_1\otimes {\cal H}_2 \otimes \cdots \otimes {\cal H}_N$, which is the tensor product of $N$ ($ \geq 2$) finite dimensional Hilbert spaces ${\cal H}_j$ of dimension $d_j\geq 2$ ($j=1,2, \cdots , N$); and has dimension $D=d_1d_2\cdots d_N$. Given any Hamiltonian $H=H^*$ acting on ${\cal H}$, the thermal equilibrium (or Gibbs) state $\rho_T$ for  temperature $T$ is $\rho_T=  \exp ( -H/T)/tr ( \exp (-H/T))$. Since the quantum system is assumed to be finite, negative temperatures are possible, but they will be disregarded here for reasons of economy although the effects described below are also present for negative temperatures.\\
Since the system has components, the question of the entanglement of the Gibbs-states arises. We recall that an arbitrary  state $\omega$ (mixed or pure) of the composite system is said to be {\em separable} or {\em unentangled} if it can be written as a mixture (finite convex sum) of pure product states.
If this is not the case, the state is said to be {\em entangled}.\\
At  present there are finite algorithms deciding whether a given state is entangled or not only for two qubits ($N=2$ with $d_1=d_2=2$);  and for $N=2$ with $d_1=2$ and $d_2=3$. The entanglement studies presented in the literature have, perforce, been restricted to investigations of the entanglement of the Gibbs-state reduced to the possible two-component subsystems, or to the study of particular so called entanglement witnesses or other entanglement monotone functions \cite{Ni,ArBo,Guetal,Wa1,Wa3,Wa2,OsNi,Wa4,GlBu,To,DoDo,WuBa,FuSoWa,BoTr}. 
In section II, we present some features of the entanglement issue for thermal equilibrium states for the bipartite systems mentioned. The rest of this introduction collects basic information  which is generally valid, i.e., for any $N$ and $D$, and some of which seems to have been overlooked previously.\\
In what follows we often identify the state $\omega$ viewed as a linear positive functional acting on operators $A$ and giving their expected-value $\omega (A)$, with the density operator $\omega$ such that $\omega (A)=tr ( \omega A)$. If the Hamiltonian is a real multiple of the identity, $H=c\cdot {\bf 1}$, then $\rho_T = {\bf 1}/D$ for all $T$, or viewed as a functional, the normalized trace which we denote by $\tau$.  Since ${\bf 1}/D = ({\bf 1}/d_1)\otimes ({\bf 1}/d_2)\otimes \cdots \otimes ({\bf 1}/d_N)$ the normalized trace $\tau$ is a separable state. We assume henceforth that the Hamiltonian $H$ is not a multiple of the identity, and let $P_-$ be the spectral projection  associated with the minimal (ground-state energy) eigenvalue $s_- (H)$  whose multiplicity we denote by $m_-$. For the ground-state $\rho_0= P_-/m_-$, one has  $\lim_{T\to 0} \rho_T = \rho_0$ where the limit can be taken in various ways. As the limit of matrix elements in any orthonormal basis you wish; or $\lim_{T\to 0} tr (\mid \rho_T - \rho_0\mid )=0$ where $\mid X\mid$ denotes the absolute value of the operator $X$; etc. Moreover $\lim_{T \to \infty} \rho_T = \rho_{\infty}=\tau$ in the same sense, and the map $0 \leq T \mapsto \rho_T$ is continuous.\\
 For what follows it will be important to observe that 
the energy $U(T)=\rho_T (H)$ as a function of temperature, is a monotone increasing continuous function with $\lim_{T\to 0}U(T)=U(0)=s_- (H)$ and $\lim_{T \to \infty} U(T)=\tau (H)=tr(H)/D$.

\subsection{Various {\textquotedblleft critical\textquotedblright}  temperatures\label{crit}}
A critical or threshold temperature above which entanglement is impossible has been observed in all cited studies. It is shown in \cite{Ra} that for every Hamiltonian there exists a finite critical temperature $T_S$ which satisfies: (i) $0 \leq T_S < \infty$; (ii) $\rho_T $ is separable for every $T \in [T_S, \infty ]$; and (iii) for every $0\leq T < T_S$  there are entangled thermal states with temperatures in the interval $[T, T_S )$.
We call $T_S$ the {\em lower separability temperature}.  
Even if $H$ is not a multiple of the identity, it could still be trivial in the sense that it contains no interactions whatsoever between the component subsystems, $H=\sum_{n=1}^N H^{(n)}$, $H^{(n)}$ acting only on the $n$-th component.  In this case $\rho_T=\rho_T^{(1)}\otimes \rho_T^{(2)}\otimes \cdots \otimes\rho_T^{(N)}$ is a product-state, hence  unentangled and  $T_S=0$.
A very poor upper bound on $T_S$ was given in \cite{Ra}; and the question of what happens below $T_S$ was posed and left open. It will be essentially answered in section II. \\
It is important to stress the fact that the convex set of separable states is closed, and this implies that the  set of non-negative temperatures $T$ for which $\rho_T$ is unentangled, is closed. Accordingly, the non-negative temperatures $T$ for which $\rho_T$ is entangled is an open set in $[0, T_S)$. \\
By their very definition, entanglement witnesses or entanglement monotone functions will always provide temperatures  for which the thermal states are entangled and these temperatures are lower bounds on $T_S$. A particularly simple witness is the energy itself as observed in  \cite{To,DoDo,WuBa}. Let
$\eta = \inf \{ \omega (H) : \; \mbox{$\omega$ is separable}\}$, 
that is the lowest energy expectation value obtainable with an unentangled state. The infimum is assumed, and since the map $\omega \mapsto \omega (H)$ is convex-linear, it can be taken over the pure product-states.\\
  If for some $T_1\geq 0$ we have $\rho_{T_1} ( H) < \eta$, then $\rho_{T_1}$ is entangled. By the monotone increase of $U(T)$ and the intermediate-value theorem there is a unique $T_H > T_1$ such that  $U(T_H)= \eta $ and all Gibbs states with temperatures in $[0,T_H)$ are entangled (in particular $\rho_0$ is entangled). In \cite{DoDo}, $T_H$ is denoted by $T_E$ and called the entanglement-gap temperature. It follows that $T_H \leq T_S$. It does not follow that Gibbs states with temperatures (immediately) above $T_H$ are necessarily separable. Although $T_H> 0$ does indeed signal the presence of thermal entanglement at low enough temperatures, its importance should not be overrated. The correct critical value is:
$ T_E = \inf \{ T \geq 0 : \; \rho_T \mbox{ is separable}\}$; for which one can prove, as in \cite{Ra}, that: (i) $0 \leq T_E \leq T_S$, and $\rho_{T_E}$ is separable; and (ii) $T_E >0$ if and only if $\rho_0$ is entangled, and in this case all thermal states with temperatures in $[0,T_E)$ are entangled.
Alternatively, $T_E$ could be defined as  the greatest temperature such that all Gibbs-states with temperatures below it are entangled; we call $T_E$ the {\em upper entanglement temperature}.  Obviously, $T_H\leq T_E$, but one should expect that,  in general,  $T_H$ can be a rather poor lower bound on $T_E$. The following example should serve as illustration.\\
Consider two qubits; and suppose the minimal energy $s_-(H)$ of your Hamiltonian is doubly degenerate with ground-state vectors $\psi_1=\alpha \otimes\alpha$, and $\psi_2= (\alpha\otimes\beta +\beta\otimes \alpha )/\sqrt{2}$ where $\alpha$ (resp. $\beta$) is an eigenstate of $\sigma_3$ to the eigenvalue $1$ (resp. $-1$) for one qubit. Then the ground-state is $\rho_0 =(1/2)\mid \psi_1\rangle \langle \psi_1 \mid  + (1/2)\mid \psi_2\rangle \langle \psi_2 \mid$, and it is entangled (the partial transpose of $\rho_0$ has a negative eigenvalue); $\eta=s_-(H)= \langle \psi_1,H\psi_1\rangle$ and thus $T_H=0$; but $T_E > 0$.

\subsection{The {\textquotedblleft more mixed than\textquotedblright}    ordering of thermal states. Upper bounds on $T_S$\label{succ}}
The theory of the \textquotedblleft more mixed than\textquotedblright partial ordering of states of a quantum system, \cite{AlUh}, can be put to use in the discussion of entanglement of Gibbs states. It is a result of Wehrl and Uhlmann (cf. Refs. \cite{AlUh,OhPe}), that
$0\leq T < T' \leq \infty$ implies $F(\rho_{T}) \leq F( \rho_{T'})$,
for every unitarily invariant, concave, continuous real-valued functional $F$ defined on states. 
It is shown in \cite{Ra}, that for every such functional $F$, for which $F(\omega )=F( \tau )$ implies $\omega =\tau$, there is a critical constant $C_F< F( \tau )$ such that:
(i) If the state $\omega$ satisfies $F( \omega )\geq C_F$ then $\omega$ is separable; and (ii) For every possible value $C$ of $F$ below $C_F$ there is an entangled state $\phi$ with $F( \phi )=C$.
There is an analogous version of this for unitarily invariant, convex, continuous real-valued functionals. Thus, every unitarily invariant, continuous real-valued functional which isolates $\tau$ and is either convex or concave, acts as a separability detector and can be used to obtain an upper bound on $T_S$. Indeed, take such a concave separability detector $F$. Then,  $T \mapsto F( \rho_T)$ is a non-decreasing continuous function for which $\lim_{T \to \infty} F( \rho_T)=F( \tau ) > C_F$. By the intermediate-value theorem there is $T_F< \infty$ such that $F( \rho_{T_F})=C_F$ and all Gibbs states with temperatures in $[T_F, \infty ]$ are unentangled. It follows that $T_S \leq T_F$.\\

\subsection{Thermal entanglement scenarios}
Many of the available studies, e.g. \cite{Ni,ArBo,Wa2,Wa3,OsNi,To,FuSoWa}, observe that the bipartite subsystem entanglement of   thermal states need not be monotone in temperature.\\ 
The above results allow one to distinguish various entanglement scenarios. The uninteresting scenario occurs when $T_E=T_S=0$ as happens when the Hamiltonians present no interactions whatsoever. But this scenario is possible even when interactions are present, cf. Section II.\\
The next scenario is that in which the ground-state is separable, i.e., $T_E=0$, but $T_S >0$. Then as temperature increases away from zero, separability is lost at some temperature $0 <T_1 < T_S$. For temperatures in the {\textquotedblleft separable segment\textquotedblright} $[0,T_1]$  all Gibbs states  are unentangled, and the segment $(T_1 , T_S)$ contains temperatures for which the corresponding thermal states
are entangled. The possibility arises for  various closed {\textquotedblleft separable segments\textquotedblright} ($[0,T_1], [T_2, T_3]$, $\cdots [T_n,T_{n+1}]$)  alternating with open {\textquotedblleft entanglement segments\textquotedblright} ($(T_1,T_2)$, $\cdots (T_{n+1},T_S)$). We will present examples for this {\textquotedblleft abnormal\textquotedblright} scenario in section II.\\

The other scenarios occur when the ground-state $\rho_0$ is entangled, i.e., $T_E >0$. The normal case is $T_E=T_S$ and this is what has been observed in most  studies we know of. Although, entanglement   usually decreases with increasing temperature, one can have non-monotonous behaviour, cf. section II. 
The abnormal scenario is $0<T_E < T_S$, and then there is a temperature $T_1$ with $T_E \leq T_1 < T_S$ such that for $T \in [T_E, T_1]$, $\rho_T$ is unentangled but there are temperatures $T'\in (T_1, T_S)$ for which $\rho_{T'}$ is entangled. Again the way is open for closed separable segments alternating with open entanglement segments. Examples of this behaviour are given in section II.\\

\subsection{The modulus of separability}
Consider any state $\omega$ of the composite system and consider the segment joining the normalized trace $\tau$ to $\omega$, i.e., $\omega_t = t\cdot \omega +(1-t)\cdot \tau$ with $0\leq t\leq 1$. Since $\omega_0=\tau$ is separable, one will ask in case $\omega_1=\omega$ is entangled,   when as $t$ increases does one lose separability? This question has been analyzed by many authors, notably by  \.{Z}yczkowski, 
Horodecki, Sanpera, and Lewenstein, \cite{Zy}, who develop it to obtain a method of estimating the ``size'' of the separable states; and by Vidal , and Tarrach, \cite{ViTa}, who give a virtually complete treatment of the problem. In Ref. \cite{Ra}, the modulus of separability of $\omega$ was defined as
$\ell ( \omega ) = \sup \{ 0 \leq t \leq 1: \; \omega_t \mbox{ is separable }\}$, whereas the quantity considered by Vidal and Tarrach is $R( \omega \mid\mid \tau )= (\ell ( \omega ))^{-1} -1$ and called by them {\em the  random robustness of entanglement}. Here we only need to observe that $0 < \ell ( \omega )\leq 1 $, with $\ell (\omega )=1$ if and only if $\omega $ is unentangled. Moreover, the upper-semicontinuity property of $\ell$ obtained in \cite{Ra}, guarantees that the map $T \mapsto \ell ( \rho_T )$ is continuous. This in turn, proves the claims made in \S\ref{crit} about the sets of temperatures where the Gibbs state is separable, respectively entangled.

\section{Thermal entanglement in qubit/qubit and qubit/qutrit systems}

Vidal and Tarrach, \cite{ViTa} have computed the modulus of separability for a qubit/qubit ($D=4$) or a qubit/qutrit ($D=6$); their beautiful formula is
\[ \ell ( \omega ) = \frac{1}{1+D \, \mid\min \{ \lambda (\omega ), 0 \}\mid}\;,\]
where $\lambda (\omega )$ is the minimal eigenvalue of the partial-transpose of $\omega$ with respect to the qubit. The plots which we will exhibit show $T \mapsto \ell ( \rho_T )$ for selected Hamiltonians which exemplify the distinct scenarios.\\
The critical values (or good upper and lower bounds) are known for some functionals (e.g. partial ordered eigenvalue sums, von Neumann entropy) from Ref. \cite{Ra} and can thus be used to provide upper bounds on $T_S$ as described in \S\ref{succ}. The map $\omega \mapsto tr ( \omega^2)$ is strictly convex (cf. Ref. \cite{OhPe}, p. 47), unitarily invariant and is easily seen to isolate $\tau$. Furthermore, the critical value for the trace of the square is known to be $1/3$ for two qubits and it lies between $1/5$ and $7/32$ for a qubit/qutrit system (\cite{Ra}). It turns out to be a rather useful separability detector because it is easy to calculate. Although it often provides a rather poor upper bound on $T_S$, in all the examples to be presented below it gave better bounds than those obtained using the partial eigenvalue sums, or the von Neumann entropy.
 We denote by  $T_*$,  the numerically obtained value of $T_F$ --recall \S\ref{succ}--  for $F$ equals minus the trace of the square (using the bound $1/5$ for the qubit/qutrit case).
We have not invested a lot of effort in the computation of $\eta$ of \S\ref{crit}. The temperature value $T_H$ obtained from the numerical, possibly inexact, value of  $\eta$ need not  necessarily be a lower bound for $T_E$ or even $T_S$; but in our experiments this never happened. The eigenvalues of the Hamiltonian counting multiplicities are given as a row-vector ${\bf h}$. Since in our definition of $\rho_T$ we have incorporated Boltzmann's constant in the Hamiltonian, the components of ${\bf h}$ have the same dimension as the temperature. Since thermal equilibrium states are invariant with respect to addition of a multiple of the identity to the Hamiltonian, we choose $s_-(H)=0$. The eigenvector to the $j$-th eigenvalue $h_j$ is listed as a row vector  $ e_j$,  where the coordinates are with respect to the canonical orthonormal tensor-product basis built from the orthonormal basis $\{(1,0), (0,1)\}$ of ${\mathbb C}^2$, and $\{(1,0,0),(0,1,0),(0,0,1)\}$ of ${\mathbb C}^3$.
\\

The Hamiltonians to be presented are specifically chosen to exhibit transitions from entanglement to separability below $T_S$, that is in the interval $[0, T_S )$. The general idea is, obviously, to choose  the eigenvector  associated with the non-degenerate ground-state energy to be either separable or entangled and then the eigenvector associated to the first excited state to be, correspondingly, either entangled or separable.
  
\subsection{qubit/qubit}
The minimal value of the separability modulus for two qubits is $1/3$.\\
Figure 1, shows the $0=T_E < T_S$ scenario with a separable segment $[0,T_1]$ followed by an entanglement segment $(T_1,T_S)$.  Using the very same eigenvectors as those of Fig. 1, but with ${\bf h} =(0,1.5,2,3)$, $\rho_T$ is unentangled for every $T\geq 0$, that is $T_S=0$.\\
Figure 2 shows two instances of the standard normal scenario ($T_H <$) $T_E=T_S$.\\
In the qubit/qubit system, we have not found the scenario where $0<T_E<T_S$  (cf. Figure 3 for a qubit/qutrit).

\subsection{qubit/qutrit}
The minimal value of the separability modulus for a qubit/qutrit systems is $1/4$.\\
Figure 3 shows the $0<T_E<T_S$ scenario; the entanglement segment $[0,T_E)$ is followed by a separability segment $[T_E,T_1]$ and a second entanglement segment $(T_1,T_S)$. In this an further examples not  shown here, we have found that once a region of the ${\bf h}$-space is found where the pertinent scenario is present, the values $T_E,T_1,T_S$ are quite stable with respect to changes in ${\bf h}$ which are large with respect the temperature values. The inset of Figure 3 shows this effect. Moreover, the same eigenvectors used in Fig. 3, but with ${\bf h}=(1.7,0,1.75,2,3,4)$ give $T_E=T_S=0.699$.\\
Figure 4 shows the $0=T_E< T_S$ scenario (cf. Figure 1 for the two qubit system) with  four transitions: $[0,T_1]$, and $[T_2,T_3]$ are separable segments alternating with entanglement segments $(T_1,T_2)$ and $(T_3, T_S)$. 
 
\section{Concluding remarks} 

We have introduced two characteristic temperatures $T_E$ and $T_S$ which  organize the entanglement behaviour of the thermal state associated to any Hamiltonian of an arbitrary composite system. For qubit/qubit and qubit/qutrit systems, we have exemplified the possibility of various transitions from entanglement to separability as temperature increases from zero to $T_S$, above which entanglement is impossible. One could expect that the features found here will persist and be enhanced as $N$ or $D$ increase (although at present there is no manageable criterion to decide when a given state is entangled or not).  Thus, in general, there will be many separable temperature segments alternating with entangled ones for multipartite systems of higher dimensions.\\

The main motivations for the present study came from certain problems posed in Ref. \cite{Ra}. From that point of view, the main conclusion to be drawn from our findings here are theoretical and  concern the results briefly mentioned in \S\ref{succ}. It was asked in Ref. \cite{Ra}:  given a separable state $\omega$ which is not pure, does there exist a unitarily invariant, concave, continuous real-valued functional $F$ defined for the states of  the composite system which isolates the trace and  such that $F( \omega )\geq C_F$?  The answer given here is definitely no! Take any unentangled thermal state $\rho_{T_1}$
such that for some $T_2 >T_1$ the Gibbs state $\rho_{T_2}$ is entangled. Then there cannot exist an $F$ with $F( \rho_{T_1})\geq C_F$ because by Wehrl's result (cf. \S\ref{succ}), $F( \rho_{T_2})\geq C_F$ and by \cite{Ra}, the separability of $\rho_{T_2}$ would follow.\\

\newpage
\begin{figure}
\begin{center}
\psfig{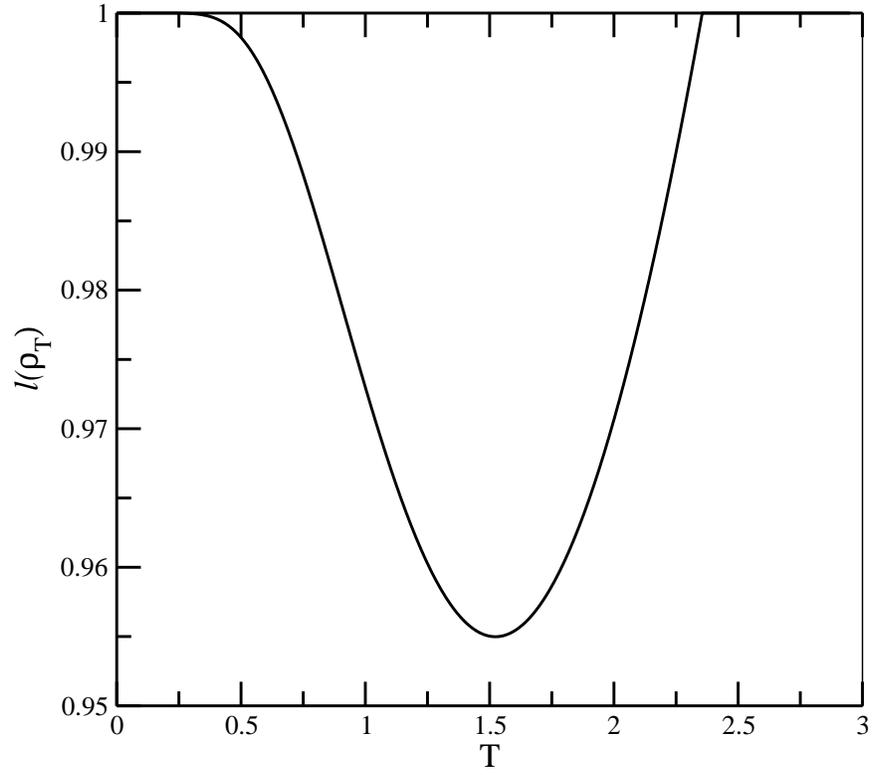}
\end{center}
\caption{$e_1=(1,0,0,0)$, $e_2=(0,x,y,0)$, $e_3=(0,x,-x^2/y,z/y)$,
$e_4=(0,z,-xz/y,-x/y)$, where $x=0.5$, $y = \sqrt{1-x^2}$ and $z =
\sqrt{1-2x^2}$; ${\bf h}=(0,1.5,7,8)$. $T_H=T_E=0$, $T_1=0.159$, $T_S=2.356$,
and $T_*=5.40$. }
\end{figure}

\newpage
\begin{figure}
\begin{center}
\psfig{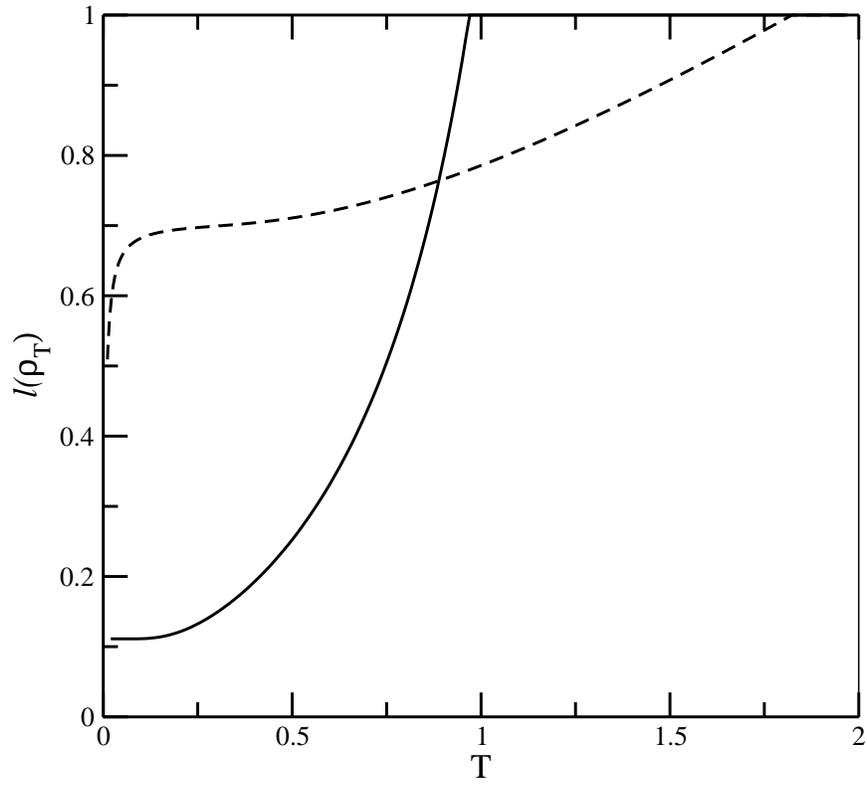}
\end{center}
\caption{Solid line:  $e_1=(1/\sqrt{2},0,0,1/\sqrt{2})$,
$e_2=(1/\sqrt{2},0,0,-1/\sqrt{2})$, $e_3=(0,1/\sqrt{2},1/\sqrt{2},0)$,
$e_4=(0,1/\sqrt{2},-1/\sqrt{2},0)$; ${\bf h}=(0.75,0,0.75,2)$. $T_H=0.73$,
$T_E=T_S=0.97$,  and $T_*=1.04$. Dashed line: $e_1=(1,0,0,0)$,
$e_2=(0,1/\sqrt{2},1/\sqrt{2},0)$, $e_3=(0,1/\sqrt{2},-1/\sqrt{2},0)$,
$e_4=(0,0,0,1)$; ${\bf h}=(0.01,2,0,4)$. $T_H=0.377$, $T_E=T_S=1.823$,  and
$T_*=2.181$}
\end{figure}

\newpage
\begin{figure}
\begin{center}
\psfig{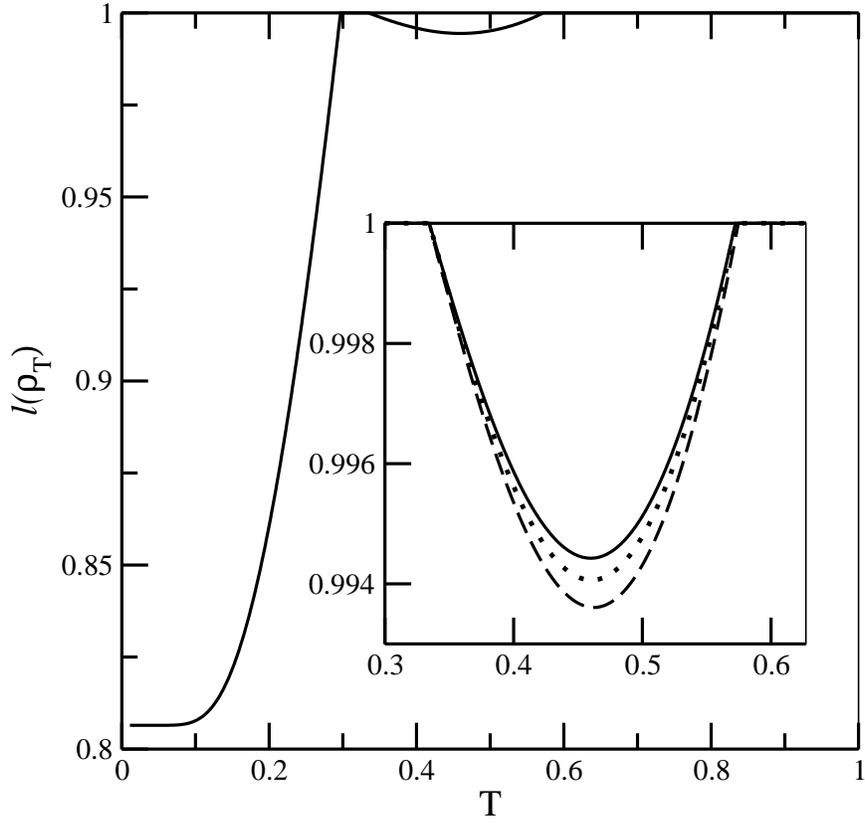}
\end{center}
\caption{$e_1=(1,0,0,0,0,0)$, $e_2=(0,0,x,0,x,y)$,
$e_3=(0,0,1/\sqrt{2},0,-1/\sqrt{2},0)$,
$e_4=(0,1/\sqrt{2},0,1/\sqrt{2},0,0)$,
$e_5=(0,1/\sqrt{2},0,-1/\sqrt{2},0,0)$,
$e_6=(0,0,y/\sqrt{2},0,y/\sqrt{2},-x\sqrt{2})$, where $x=0.2$ and
$y=\sqrt{1-2x^2}$; ${\bf h}=(0.75,0,0.75,2,3,4)$. $T_H=0.13$, $T_E=0.296$,
$T_1=0.334 $, $T_S=0.571$,  and $T_*=2.76$. Inset: using the same eigenvectors
and eigenvalues except, from top to bottom, $h_1=0.75, 1$ and $1.5$}
\end{figure}

\newpage
\begin{figure}
\begin{center}
\psfig{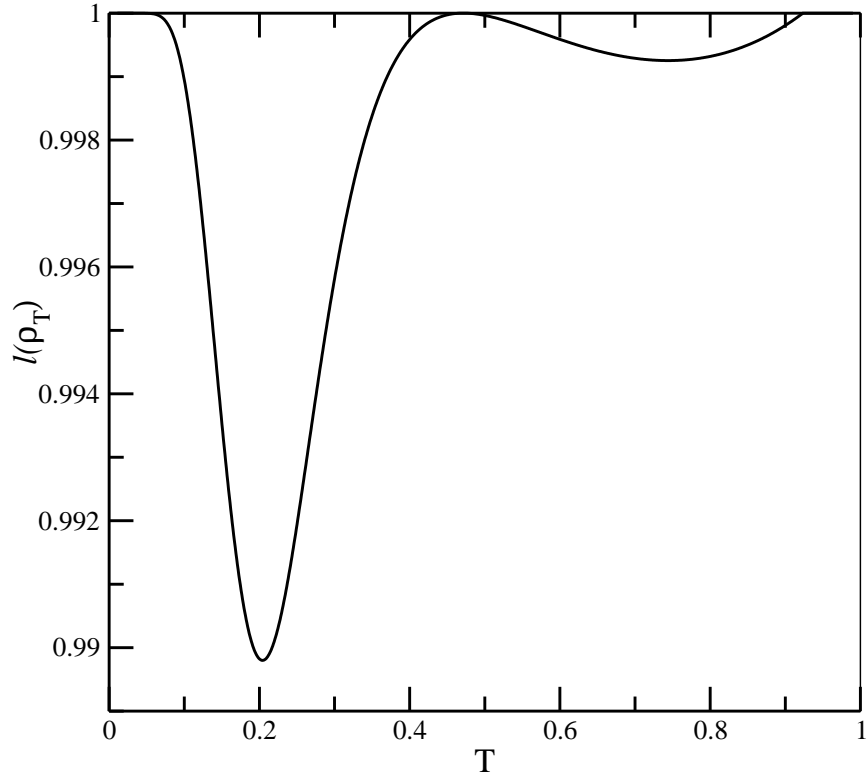}
\end{center}
\caption{$e_1=(1,0,0,0,0,0)$, $e_2=\frac12(0,1,0,1,1,1)$,
$e_3=(0,0,1,0,0,0)$, $e_4=\frac12(0,1,0,1,-1,-1)$,
$e_5=\frac12(0,1,0,-1,1,-1)$, $e_6=\frac12(0,-1,0,1,1,-1)$; ${\bf
h}=(0,0.7,7,0.9,1,1.5)$. $T_H=T_E=0$, $T_1=0.0355 $, $T_2=0.467 $,
$T_3=0.476$, $T_S=0.923$,  and $T_*=2.645$.}
\end{figure}

\end{document}